# Citation personal display: A case study of personal websites by physicists in 11 well-known universities

Xingchen Li, Qiang Wu* and Nan Zhang

*School of Management, University of Science and Technology of China,*

*96 Jinzhai Road, Hefei 230026, China*

[*Corresponding author. Qiang Wu can be contacted at: qiangwu@ustc.edu.cn]
(Dated: February 25, 2017)

**Abstract**

**Purpose** – This paper aims to investigate the extent to which researchers display citation, and wants to examine whether there are researcher differences in citation personal display at the level of university, country, and academic rank.

**Design/methodology/approach** – Physicists in 11 well-known universities in USA, Britain, and China were chosen as the object of study. It was manually identified if physicists had mentioned citation counts, citation-based indices, or a link to Google Scholar Citations (GSC) on the personal websites. A chi-square test is constructed to test researcher differences in citation personal display.

**Findings** – Results showed that the overall proportion of citation personal display is not high (14.8%), with 129 of 870 physicists displaying citation. And physicists from different well-known universities indeed had a significant difference in citation personal display. Moreover, at the national level, it was noticed that physicists in well-known Chinese universities had the highest level of citation personal display, followed by Britain and the USA. Further, this study also found that researchers who had the academic rank of professor had the highest citation personal display. In addition, the differences in h-index personal display by university, country or academic rank were analyzed, and the results showed that they were not statistically significant.

**Originality/value** – This is the first study to investigatehow widely researchers provide citation-based information on personal websites.

# 1. Introduction

Citation is a behavior in which one cites researchers' theories or research findings. To illuminate the citing behavior, two complete theories have been explored (Bornmann and Daniel, 2008). One is the normative theory (Merton, 1973; Merton, 1988) and the other is the social constructivist view (Knorr-Cetina, 1981; Gilbert, 1977; Collins, 2004). The normative theory argues that scientists give recognition, reputation, or credit to authors whose works are cited, while the constructive approach states that scholars have complex citing motivations and try to persuade the readers to accept their findings because of the cited works. In practice, many scientists have studied citing behavior from different angles. Garfield (1962) listed possible citing motivations which were classified into fifteen types, such as paying homage to pioneers, giving credit for related work, and correcting one's own work (Garfield, 1962; Bornmann and Daniel, 2008). Vinkler (1987) gave an excellent summary of citing motivations, which were divided into two major groups: professional motivations (i.e. acknowledging or criticizing theoretical and practical content of the cited papers) and connectional citing behavior (i.e. establishing relations with scientific community). Vinkler also found that citations were mainly influenced by professional motivations and hence citations could be reliable for evaluative bibliometric analyses.

Kostoff (1998) studied the possible uses of citations by focusing the applicability of citation analysis as aquality measure, and considered that citations could help readers to save much time by serving as a bookmark which provides a condensed reference to a huge amount of information content. Not only that, the citations' roles also include intellectual heritage, impact tracking, and self-serving purposes. It is an undeniable fact that citation analysis should not be used as a stand-alone measure of research quality or impact. Peer review also plays a significant role in evaluating scientists' research quality or impact and it may offer more plentiful information. However, peer review seems subjective. Compared to peer review, the significant advantage of citations is that they provide an objective measure of the impact and so citation analysis makes the judgement more objective (Garfield, 1979; Schmoch and Schubert, 2007). The feature of objectivity makes citations a useful yardstick to evaluate research impact in bibliometrics. Lehmann et al. (2006) mentioned that the number of citations is a widespread and useful measure of individual scientist quality.

Sidney Redner agreed "it would be useful to have an objective criterion for election to bodies such as the US National Academy of Sciences (NAS) or Britain's Royal Society" (Ball, 2005), which further indicates that the number of citations, an objective criterion, has significant value. Besides, Meho (2007) pointed that citation data have been used by many governments, funding agencies, and tenure and promotion committees to assess the quality of a researcher's work. Therefore, the presence of citation information is more likely to transform the scientist's qualitative evaluation into quantitative evaluation, and this can also promote the fairness of the evaluation.

Noting increasing concern of measuring research achievement based on citation, Hirsh (2005) presented his h-index/Hirsch-index which is used to quantify an individual researcher's scientific output. As Hirsh claimed, the h-index is a transparent and impartial indicator. Right after the h-index was proposed, other citation-based indices appeared, such as the g-index (Egghe, 2006), r-index (Jin et al., 2007), and w-index (Wu, 2010). All of the indices mentioned here can be adopted to evaluate scientists' achievements, and all of them have the characteristic of objectivity which leads researchers to consider them convincing and acceptable. Studies have shown that the h-index/Hirsch-index has been gradually accepted by many institutions and scientists, as illustrated by its acceptance by the journals Nature (Ball, 2005) and Science (Anonymous, 2005). Lehmann et al. (2006) argued that some citation-based measures could be helpful, provided there is a sufficient quality of data. Besides, Ball (2007) drew a conclusion that the h-index did have the ability to distinguish good scientists and is widely used informally. However, not all scientists agree that the h-index or other citation-based indices can be used to evaluate a scientist's achievement or quality of publications. Döring (2007) considered citation-based indices to be insufficient to evaluate the quality of publications, regardless of its statistical reliability. It would be interesting to know how widely these citation-based indices are mentioned, but no previous study had investigated this issue, which led to the investigation of the current paper. The goal of this research is to study the extent to which researchers display citation-based indices on their personal websites.

The rapid development of science and technology has made the Internet become the main source of information. Information on the Internet takes many forms: common email, CV, open access repositories, personal websites, blogs, etc. In fact, some studies have shown that researchers are more and more dependent on electronic resources(Herring, 2002). The personal website, as a form that can propagate information, has been gradually noticed and analyzed by researchers (Ajiferuke and Wolfram, 2004; Barjak et al., 2007; Más-Bleda and Aguillo, 2013; Kousha and Thelwall, 2014; Más-Bleda et al., 2014a; Más-Bleda et al., 2014b). Researchers who have personal websites usually include their CVs in the personal websites. The CV is already seen as a good place to include all kinds of information, such as a researcher's education, position, grants, and awards. This information could reflect an individual's value from different aspects and help scientists to accumulate their credit (Latour and Woolgar, 1986). With regard to scientists, Barjak et al. (2007) pointed out that many scientists publish scientific research on their personal websites or research group websites. After studying the personal websites of European highly-cited researchers, Más-Bleda and Aguillo (2013) pointed out that personal websites of scientists could be an excellent pathway to disseminate their research findings, and they also believed that one could evaluate a scientist based on his/her personal website. Thus, it can be seen that such a website is an extremely useful place where valuable information is available for judging research quality. Consider, for example, Tom Abel, who is a physicist working at Stanford University and gives his research area, group information, publications, press, citations (12100), h-index (57), i10 index (97) etc. on his personal website (http://www.tomabel.org/Home/ Welcome. html).

Despite a number of works that have extended the study of personal websites, none of these works has focused on the degree to which scientists display citation counts or citation-based indices on their personal websites. In this paper, we aim to fill this gap by manually identifying whether scientists display citations or citation-based indices. If the researchers have a relatively high proportion of citation personal display (for example, 80% display their citation counts or citation-based indices on their personal websites), then we have evidence that the scientists being evaluated consider citation-based indices (e.g. h-index, g-index) as important indicators to evaluate research contributions.

## 2. Citation personal display

Here, the term "citation personal display", coined by the authors, means researchers mentioning their citation counts or citation-based indices on their own personal websites. We believe that citation personal display embodies three aspects. Firstly, it may represent a kind of recognition of index validity, that is to say researchers admit that citations can be a yardstick that can weigh their research achievements. Secondly, because the number of citations or the numerical value of citation-based indices has a magnitude, the scientists also believe that the magnitudes of the indices are useful indicators to distinguish levels of researchers' academic achievement. Thus, those who put citation data on their websites acknowledge the value of citation in highlighting their own achievements. Thirdly, the scientists are providing data for quantitative assessment, knowing that citation or citation-based indices cause hiring decisions, funding evaluations, and promotions to be objective, and thereby researchers are implicitly approving of using citation data for these purposes.

In this study, we consider researcher citation personal display to exist as long as one or more of the following conditions ismet on the researcher's personal website:

(1) Citation counts are mentioned;
(2) The researcher provides his/her own h-index or other citation-based indices (e.g. g-index; r-index);
(3) A link is given to Google Scholar Citations (GSC) where the researcher's total citations and h-index can be obtained.

We also coin another new term "h-index personal display", which means a researcher displays the h-index or provides a link to Google Scholar Citations (GSC) on his/her own personal website.

Although there are various social media applications that provide citation-related information, we consider a link to Google Scholar Citations as a researcher citation personal display for the following reasons. Firstly, Google Scholar Citations is a free search engine and the major function of Google Scholar Citations is to provide citation-based information (Yang and Meho, 2006; Meho and Yang, 2007; Levine-Clark and Gil, 2008). Secondly, although there are other profiling services that use h-index or citations, like ResearchGate, most of them are used to provide communication platforms for researchers rather than authoritative citations sources. In the academic community, there are three widely-used databases, namely Scopus, Web of Science, and Google Scholar, which are viewed as main sources for citation analysis. Considering that among these only Google Scholar is free for everyone, it is convenient for scientists to access a page of Google Scholar Citations. Furthermore, no scientist provided links to Scopus and Web of Science in our study because they are not easy to obtain. Hence, this study pays attention to links to Google Scholar Citations.

Of course, there remains one case to consider, that is, a researcher who has a public Google Scholar profile but does not provide its link on the personal website. For this case, we think that those researchers do not recognize citations sourcing from Google Scholar Citations, or do not want to directly display citations on the personal website. Since citation itself contains complicated facets and there are different perspectives on viewing citations, this work only focuses those who have definite attitudes towards displaying citations. Hence, researchers who have but do not give a link to a Google Scholar profile on their personal websites will be deemed "non-citation displayers".

One thing we want to point out here is that Microsoft Academic Search is also considered as an important free citation-based academic search engine, which offers some citation-based information like Google Scholar Citations (Ortega and Aguillo, 2014). Since Microsoft Academic Search was started only in 2009, it lags behind Google Scholar Citations and its usage is less widespread. Among the websites examined in this study, almost none gives a Microsoft Academic Search profile. Therefore, this work mainly focuses on researchers who have a link to Google Scholar Citations on their personal websites. In the later work, there is no doubt that these links like Microsoft Academic Search should be regarded as citation personal display if a significant number of researchers provide such profiles on their personal websites.

## 3. Research questions

We selected eleven well-known universities as the object of study, and analyzed the extent of researcher citation personal display on the basis of researchers' personal websites. Based on the above settings, this paper primarily studies the following questions:

(1) Do the researchers discussed in this study have a high degree of citation personal display?

(2) Do researchers in eleven well-known universities have varying extents of citation personal display?

(3) Do researchers who are in different countries vary in their citation personal display?

(4) Does the extent of citation personal display vary from academic rank to academic rank?

## 4. Methods

### 4.1 The constitution of personal website

The personal website, as a serviceable platform for researchers publishing various sorts of information, can have different forms ranging from institutional pages to non-institutional websites. The common forms of non-institutional website are: thematic repositories (ArXiv, RePEc), social websites (Facebook, LinkedIn), and scholarly databases (Google Scholar, ResearchGate, Microsoft Academic Search, CiteULike). Although these channels can offer web space to researchers, their credibility may be weaker when compared to institutional personal websites which researchers maintain. Not only that, the non-institutional personal website does not seem widespread due to its comparative informality. Although its recognition and development trend may change in the future, the institutional personal website is popular in researchers at present. An additional advantage of institutional personal website is that it can be found easily from the corresponding institution's official website. Furthermore, a researcher's institutional personal website can provide comprehensive information which often is hard to obtain, such asthe complete CV, research projects, conference presentations (Más-Bleda and Aguillo, 2013). Hence, an institutional personal website can be a great channel for evaluating researchers.

In this study, the scope of a personal website only focused on institutional personal websites, that is, a website hosted on a faculty website. More specifically, as long as a researcher appears on a university website, department homepage, laboratory website, or research group website, the researcher is considered to have a personal website. The data of this study showed that most researchers studied had institutional personal websites, although 20 researchers in China did not. The high proportion of researchers maintaining such a personal website is attributed to the target objects discussed in this paper. All universities in this study were chosen because they are well known, and such well-known institutions all provide personal web space so potential lack of such space is not an issue.

*4.2 Selection of research objects*

Examining the rationale behind a selection process is very important for the accuracy of results. Since the h-index was proposed by Hirsch, who is a physicist, we hypothesized that researchers in physics may have a relatively high percentage of citation personal display. Furthermore, there are some studies which suggest physics scientists often use the web to disseminate information through disciplinary repositories like arXiv (Goodrum et al., 2001; Shingareva and Lizarraga-Celaya, 2012), and the chemists are less likely to use the Web than other scientists (Brown, 2007). Therefore, in this study physics was chosen as the research field in which to explore the extent of researchers' citation personal display, and physicists in well-known US universities, British universities, and Chinese universities were selected as research objects.

Here, the standard of selecting universities was different in different countries. In Britain and USA, using the Times Higher Education World University Rankings 2015-2016, we selected the top 8 universities, which include 3 famous universities in Britain (University of Oxford, University of Cambridge, and Imperial College London) and 5 famous universities in USA (California Institute of Technology, Stanford University, Massachusetts Institute of Technology, Harvard University, and Princeton University). In China, selection was done according to Chinese university rankings in physics as published in the Higher Education Evaluation of the Ministry of Education, with the top 3 universities being chosen (Peking University, Tsinghua University, and Fudan University). A list of these 11 well-known universities and abbreviations is provided in Table V of Appendix.

The data was manually collected from all physicists' personal websites. The data collection started in October 2015, and we also manually conducted a second search to guarantee the accuracy of the data, finishing the data collection in November 2015. In the process of collecting data, in order tofocus on regular university professors, we have excluded researchers who were deceased researchers, emeritus professors, postdoctoral researchers, and visiting researchers. The subjects of this paper were limited to professors, readers/associate professors, and lecturers /assistant professors. If a researcher has a personal website, the contents were analyzed in terms of four aspects: the researcher's academic rank, whether citations were mentioned, whether an h-index or other citation-based indices were provided, and whether a link to Google Scholar Citations was given.

The data collection was from 870 physicists' personal websites. Among the 870 physicists who have the personal websites, 326 (37.5%) work in China, 245 (28.2%) in Britain, and 299 (34.4%) in the USA. It is worth mentioning that all physicists who mention citation-based indices always provide the h-index, with only two physicists giving other indices like w-index. For description convenience, we only count the number mentioning h-index in the next section. To test researcher differences in citation personal displayat the level of university, country, and academic rank, the chi-square tests are constructed.

## 5. Results

*5.1 Theextent of citation personal display*

Since the focus of this study is to investigate the extent of citation personal display in 11 well-known universities, we integrate the citation-based information of these famous universities, listing the data in Table I. Using the definition of citation personal display and h-index personal display, the calculations of citation personal display and h-index personal display are as follows:

Citation personal display counts = number of directly showing citations + number of directly showing h-index + link counts of GSC - repeated ones

H-index personal display counts = number of directly showing h-index + link counts of GSC - repeated ones

Here, the number of directly displaying citations (abbreviated No. DDC) refers to researchers who directly mention their citation counts on their personal websites. Similarly, the number of directly displaying h-index (abbreviated No. DDH) also means those researchers who directly mention their h-index on the personal website. The link counts of GSC can be regarded as indirectly mentioning their citation counts or h-index on the personal website.

Table I   Distribution of information contained in personal website by university

| University | No. physicists | No. DDC | No. DDH | Link counts of GSC | Citation personal display counts |
|---|---|---|---|---|---|
| Harvard | 63 | 5 | 2 | 2 | 7 |
| MIT | 95 | 9 | 5 | 0 | 9 |
| Stanford | 47 | 7 | 6 | 0 | 7 |
| Caltech | 44 | 2 | 1 | 0 | 2 |
| Princeton | 50 | 4 | 3 | 0 | 4 |
| Oxon. | 126 | 17 | 11 | 0 | 18 |
| Cambridge | 56 | 12 | 8 | 0 | 12 |
| ICL | 63 | 8 | 6 | 1 | 9 |
| Peking | 184 | 21 | 11 | 0 | 21 |
| Tsinghua | 72 | 22 | 6 | 0 | 23 |
| Fudan | 70 | 14 | 6 | 0 | 17 |
| Sum | 870 | 121 (13.9%) | 65 (7.5%) | 3 (0.3%) | 129 (14.8%) |

Note: No. DDC: number of directly displaying citations; No. DDH: number of directly displaying h-index; No. physicists: number of physicists; GSC: Google Scholar Citations.

Table I indicates that the overall proportion of citation personal display is not high (14.8%), with 129 of 870 displaying citation. Of the 129 physicists showing citation personal display, 121 directly show citations, 65 directly show h-index, and only 3 provide the link to Google Scholar Citations. Most citation personal display spring from No. DDC and No. DDH that physicists provide directly on their personal websites. If we accept that researchers who give directly their citation counts (13.9%) or h-index (7.5%) on their personal websites may hold strong opinions about citation personal display, the distribution of citation personal display suggests that most physicists have stronger citation personal display than those who only provide links to Google Scholar Citations (0.3%).

In terms of h-index personal display, the number of h-index personal display is 68 which is the sum of No. DDH and link counts of GSC. From Table I, we find that physicists in 11 well-known universities also have a low h-index personal display (7.8%). Most of the h-index personal display comes from No. DDH, and there were only 3 links to Google Scholar Citations. In summary, the physicists discussed in this paper do not have a high proportion of citation personal display, that is, the citation counts or h-index are not widely used by physicists on their personal websites, which is counter to what was expected.

*5.2 Contrast of 11 well-known universities*

When trying to compare citation personal display in these 11 well-known universities, the proportions of citation personal display and h-index personal display were calculated (Figure 1). As Figure 1 shows, distribution of citation personal display is skewed, withseveral universities (Tsinghua, Fudan, Harvard, and Cambridge) having higher percentage. Physicists who have worked in Tsinghua have the highest proportion of citation personal display (31.9%), with 23 physicists providing their own citation counts on their personal websites, followed by Fudan (24.3%) and Cambridge (21.4%). As an exception to other universities, physicists at Caltech have the lowest citation personal display with only 2 physicists mentioning their citations, and the share of citation personal display being merely 4.5%. The share of Princeton's citation personal display is also low, with 8% physicists mentioning their citations. The rest of the universities have a moderate degree of citation personal display, all with more than 10%: Stanford (14.9%), Oxon. (14.3%), ICL (14.3%), Peking (11.4%), and Harvard (11.1%).

As it can be seen, differences in citation personal display do exist for these 11 well-known universities. But are these differences significant? This cannot be found from Figure 1. To test the significance, a chi-square test is constructed. The result indicates that this difference is statistically significant (the corresponding p< .05).

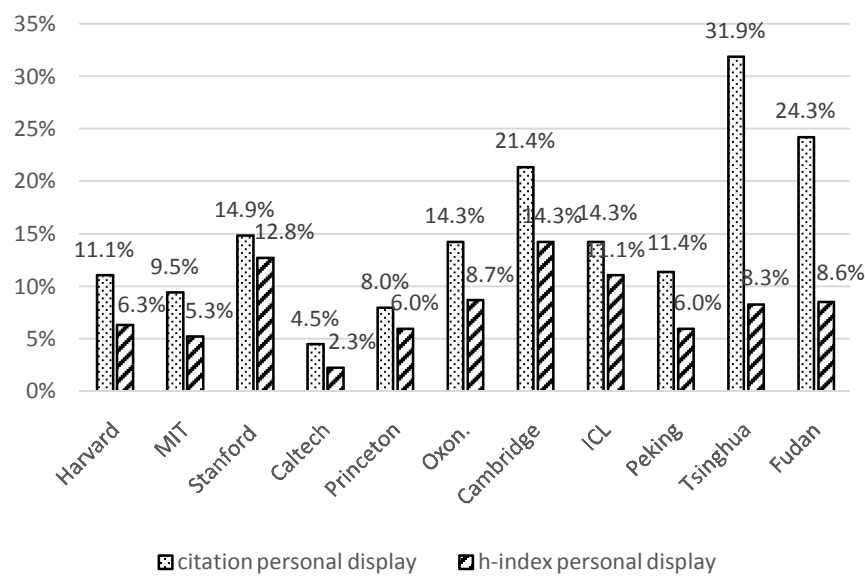

**Figure 1.** Percentages of citation personal display and h-index personal display by university

Considering h-index among 11 universities, the highest share (14.3%) of mentioning h-index is Cambridge, followed by Stanford (12.8%) and ICL (11.1%). The others are less than 10%, including Oxon. (8.7%), Fudan (8.6%), Tsinghua (8.3%), Harvard (6.3%), Princeton (6%), Peking (6%), and MIT (5.3%). The lowest proportion of displaying h-index is at Caltech, with only one physicist providing the h-index. The chi-square test was used to see whether the proportion of h-index personal display is significantly different, and the result suggests that the difference is statistically non-significant ($p > .05$).

In this paper, the above definition of citation personal display says that h-index personal display is included in citation personal display. Therefore, one would expect the trends of citation personal display should stay consistent with h-index personal display. However, the data tells a different story, and this is also the reason why we state the share of h-index personal display by university even though it is not statistically significant. For instance, from the aspect of citation personal display, Tsinghua has the highest proportion, while its share of h-index personal display (8.3%) is far less than many universities.

*5.3 Contrast by country*

In this section, we discuss the extent of physicists' citation personal display in terms of their countries. Table II summarizes the general information for physicists in well-known Chinese universities, British universities, and US universities. Then we use total number of physicists as a denominator to compute the percentage of citation personal display respectively (Figure 2).

**Table II**  Distribution of citation personal display and h-index personal display by country

| Country | Total number of physicists | Citation personal display counts | H-index personal display counts |
|---|---|---|---|
| USA | 299 | 29 | 19 |
| Britain | 245 | 39 | 26 |
| China | 326 | 61 | 23 |

Figure 2 shows the percentage of citation personal display and h-index personal display by country. It can be found that citation personal display is distributed unevenly among the countries. 61 of the 326 physicists in well-known Chinese universities (18.7%) have citation personal display. This highest percentage is not difficult to understand considering that, as shown in Table I, Tsinghua and Peking have the highest citation personal display counts. In comparison, the number of physicists in well-known British universities who provide their own citation counts is relatively small, at 15.9%. Physicists in well-known US universities have the least share of citation personal display, with 9.7% having citation personal display. The chi-square test suggests that the difference of citation personal display between countries is significant ($p < .05$), and this is also be indicated by Figure 2.

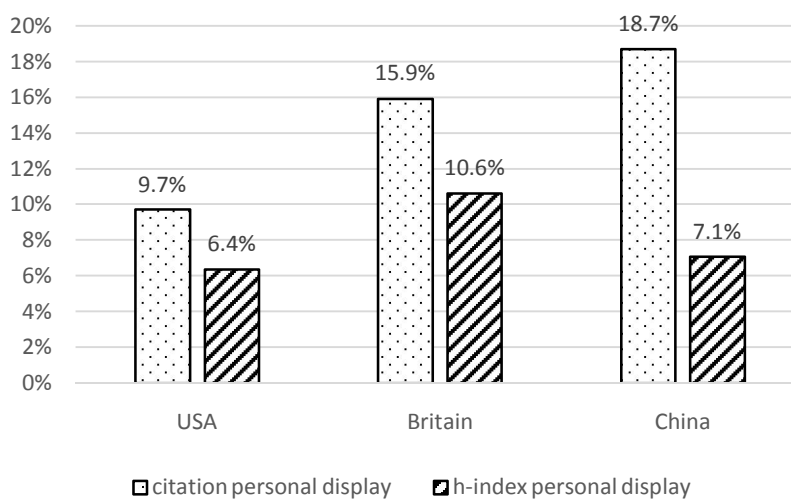

**Figure 2.** Percentages of citation personal display and h-index personal display by country

Considering h-index, is there still much discrepancy in personal display of h-index among physicists in different countries? Figure 2 also shows the percentage of h-index personal display by country. Physicists in well-known British universities have the biggest share of mentioning h-index, with 26 of 245 physicists giving their own h-index, that is, 10.6%. China ranks second, with 7.1% among the physicists discussed. Only 19 of 299 physicists in well-known US universities provide their own h-index, which is the lowest number of the three countries. This phenomenon is opposite to the above results which reveal that physicists in well-known Chinese universities have the largest share of citation personal display. However, chi-square test suggests that the discrepancy in personal display of h-index among physicists in different countries is not significant ($p > .05$).

*5.4 Contrast by academic rank*

In addition to discussing the discrepancy of citation personal display, this study also analyzes whether the researcher's academic rank is correlated with the researcher's citation personal display. The distribution of researchers' citation personal display by academic rank is listed in Table III.

Table III   Distribution of citation personal display by academic rank

| University | Professor | Reader/Associate Professor | Lecturer/Assistant Professor | Citation personal display counts |
|---|---|---|---|---|
| Harvard | 6 | 0 | 1 | 7 |
| MIT | 6 | 0 | 3 | 9 |
| Stanford | 4 | 2 | 1 | 7 |
| Caltech | 2 | 0 | 0 | 2 |
| Princeton | 3 | 1 | 0 | 4 |
| Oxon. | 11 | 3 | 4 | 18 |
| Cambridge | 7 | 0 | 5 | 12 |
| ICL | 5 | 1 | 3 | 9 |
| Peking | 13 | 8 | 0 | 21 |
| Tsinghua | 16 | 4 | 3 | 23 |
| Fudan | 11 | 5 | 1 | 17 |

As shown in Table III, physicists who are professors account for most of physicists who provide citations, and the other academic ranks (reader/associate professor and lecturer/assistant professor) have a lesser presence. Because different countries have different professional academic rank systems, we note that physicists with the academic rank "Reader" are found only at Oxon., Cambridge, and ICL. This is reflected in the raw data, not shown here. Based on guidelines for the reader academic rank, this paper views "Reader" as the same as "Associate Professor", and the academic rank "Lecturer" is also equated with "Assistant Professor".

Among 11 renowned universities, physicists working at the Caltech have the highest share of professors who have citation personal display, with 2 of 2 physicists mentioning citation (100%). Some other universities also have relatively high rates of professors who have citation personal display, such as Harvard (85.7%), Princeton (75%), and Tsinghua (69.6%). In general, the physicists who have the academic rank of reader/associate professor have more citation personal display than those who are lecturers/assistant professors. However, there are exceptions, with 5 universities (Oxford, Cambridge, MIT, ICL, and Harvard) having more citation personal display for lecturers/assistant professors than readers/associate professors.

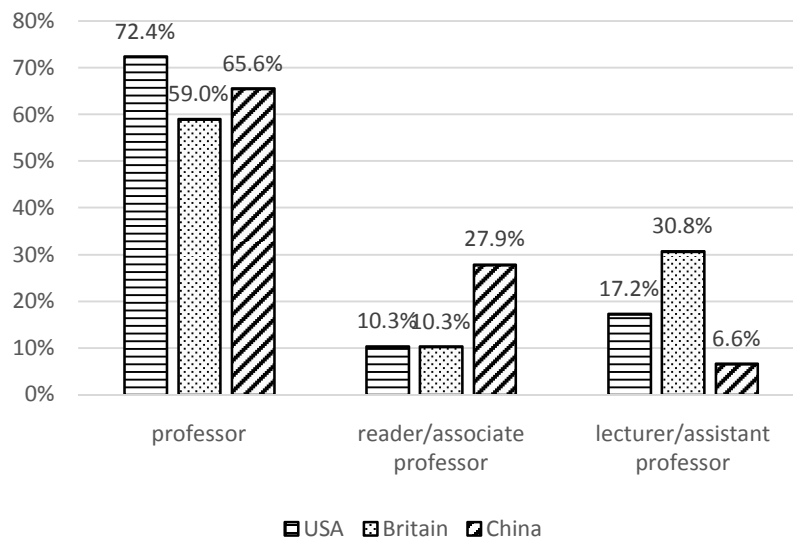

**Figure 3.**  Percentage of citation personal display by academic rank

This work also analyzes whether academic rank influences physicists' citation personal display in terms of country. Figure 3 shows the percentage of different academic rank by country. In well-known British universities, there are 23 professors of 39 physicists who have citation personal display, counting for 59%. It also has 12 lecturers who account for 30.8% of total number of citation personal display. Note that only 4 of physicists who are readers/associate professors in well-known British universities have citation personal display, which is the lowest (10.3%). The analysis also suggests that physicists who are professors in well-known US universities have more citation personal display than in the two other countries, accounting for 72.4% of citation personal display counts. In citation personal display, the percentage of assistant professors is slightly above associate professors, with 17.2% and 10.3% respectively. Similarly, for physicists in well-known Chinese universities, professors account for the majority mentioning citation, accounting for 65.6%. For China, the distribution of assistant professors and associate professors towards citation personal display is the exact opposite of the American situation; the percentage of physicists who are associate professors exceeds the percentage of physicists who are assistant professors, with 27.9% and 6.6% respectively.

To summarize, no matter whether the perspective is university or country, academic rank indeed has a relationship with physicists' citation personal display, and this correlation is statistically significant ($p < .05$). In the following sections, we will treat h-index as a start, and then discuss whether academic rank is correlated with physicists' h-index personal display. The distribution of researchers' h-index personal display by academic rank is shown in Table IV.

Table IV   Distribution of h-index personal display by academic rank

| University | Professor | Reader/Associate Professor | Lecturer/Assistant Professor | H-index personal display counts |
|---|---|---|---|---|
| Harvard | 4 | 0 | 0 | 4 |
| MIT | 3 | 0 | 2 | 5 |
| Stanford | 3 | 2 | 1 | 6 |
| Caltech | 1 | 0 | 0 | 1 |
| Princeton | 2 | 1 | 0 | 3 |
| Oxon. | 7 | 2 | 2 | 11 |
| Cambridge | 4 | 0 | 4 | 8 |
| ICL | 3 | 1 | 3 | 7 |
| Peking | 5 | 6 | 0 | 11 |
| Tsinghua | 3 | 1 | 2 | 6 |
| Fudan | 5 | 1 | 0 | 6 |

Table IV indicates that physicists who are professors prefer providing their h-index more than other physicists, but the discrepancy is not statistically significant ($p > .05$). Since the h-index itself is a kind of citation, citation personal display includes h-index personal display and therefore, in most cases, the distribution of h-index personal display is similar to Table III and Figure 3. However, there are also some differences. Taking Tsinghua as an example, its assistant professors have a relatively high share (33.3%) of h-index personal display but the corresponding share of citation personal display is low at 13%. This situation indicates that assistant professors working in Tsinghua have a high personal display of h-index compared to citation personal display.

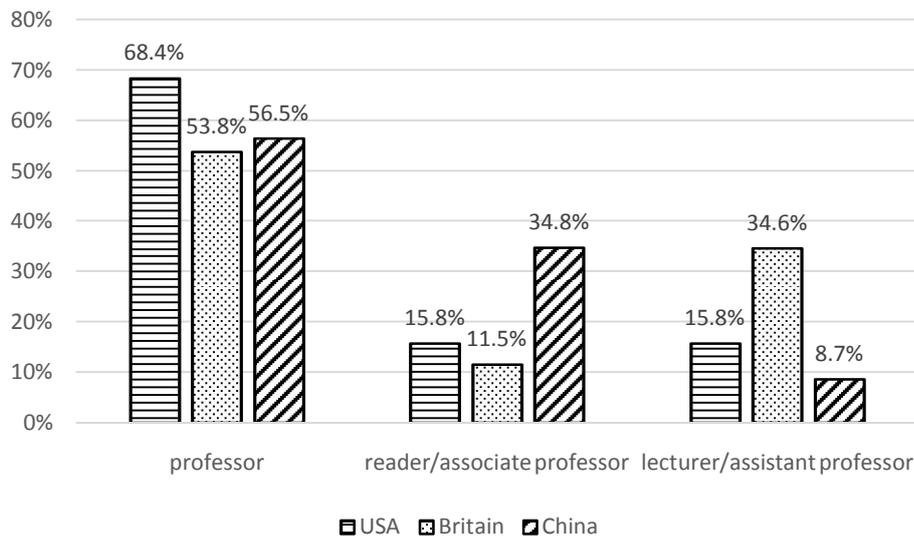

**Figure 4.** Percentage of h-index personal display by academic rank

We also compare the difference of h-index personal display from the perspective of countries, and the discrepancy is not statistically significant (p > .05). Figure 4 reflects the percentage of the physicists' h-index personal display by academic rank. Among the 3 countries analyzed in this paper, physicists who are professors in well-known US universities have the highest ratio of mentioning h-index, with 68.4% professors giving their own h-index. The remaining percentage is due to physicists who are readers/associate professors or assistant professors, counting 15.8% and 15.8% respectively. The distribution of h-index personal display in well-known Chinese universities is similar to the American situation. The difference is that h-index personal display of Chinese readers/associate professors (34.8%) is much higher than in the USA, in fact, more than double compare to the USA. Among the three countries, professors in well-known British universities have the lowest proportion of h-index personal display (53.8%), which is a little less than Chinese universities. It is worth mentioning that compared with the well-known universities in Chinaand USA, physicists who are lecturers in the British universities account for 34.6% of having h-index personal display, which is far above the figure for readers/associate professors (11.5%).

In general, the researcher's academic rank has a significant correlation with the researcher's citation personal display ($p < .05$), and no significant relationship with the researcher's h-index personal display ($p > .05$). Therefore, the higher academic rank is, the more likely the citation personal display. That is, typically a researcher who is a professor has more citation personal display than a researcher who is an associate professor/reader or an assistant professor/lecturer. There are, however, exceptions, such as lecturers in well-known British universities having more citation personal display than the associate professors/readers.

## 6. Conclusions and discussion

The main goal of this study is to explore the extent of the researcher's citation personal display by looking at whether physicists provide citation-based information on their personal websites. The personal website, as an important platform for scientists, is a promising data source. Considering the significance of the personal website, this work included manual collection of content provided on physicists' personal websites, studying physicists working in 11 well-known universities distributed in Britain, the USA, and China. Through the contrast of different universities and countries, some findings are revealed in this paper.

The study uncovers that the overall proportion of citation personal display is not high (14.8%) among physicists, but there is a significant university difference ($p < .05$) in citation personal display. Among the 11 well-known universities in question, physicists in Tsinghua have the highest citation personal display (31.9%), and the other extreme appears at the Caltech (4.5%). According to the overall situation, the citation personal display of physicists working in well-known Chinese universities is generally higher. Concerning h-index, we find that physicists at Cambridge have the highest h-index personal display, and Caltech has the lowest h-index personal display without a doubt. And yet, the difference of h-index personal display between universities is not significant ($p > .05$). The significant university differences in citation personal display reveal that physicists have different attitudes towards citations, which may be caused by university policies or influenced by peers.

Another finding of our study is that citation personal display exhibits differences between countries. On the one hand, from the point of citation personal display, physicists in well-known Chinese universities have the strongest personal display for citation among three countries, followed by Britain and the USA. The national differences in the data tell us that the degree of attention to citations or citation-based indices varies between countries, which may be caused by dissimilarities in citation culture and citing behavior in these countries. Another possibility which could explain the national differences is that every country has its own system for evaluating the quality of researchers and these different policies are likely to influence researchers' attitudes to citations. To prove the influencing factors, further studies are required, such as a questionnaire survey to the identified researchers.

Among the countries discussed in this paper, the highest proportion of citation personal display in well-known Chinese universities may reveal that Chinese physicists focus more on the quantitative aspects of evaluation when compared with their British and American counterparts. However, it must be noted that the proportion of 18.7% is not very high. This indicates that the citation-based indicators are not widely used by the physicists anywhere, although h-index has been proposed for a decade. On the other hand, from the perspective of h-index, the study suggests that physicists in well-known British universities have the highest personal display of h-index, which is exactly opposite to the results for citation personal display. However, the chi-square test suggests that the difference of h-index personal display between countries is not significant ($p > .05$).

In addition, we also find that the researcher's academic rank indeed has a significant correlation with citation personal display ($p < .05$). Researchers who are professors are more inclined to show citation, and they also have the highest share of h-index citation. However, the difference in h-index personal display by academic rank is not statistically significant (p > .05). It should be stated that there are many other factors that could contribute to why physicists display citations on personal websites besides academic rank, such as researchers' total citations and scientific age. Since the main objectives of this work are to introduce the concept of "citation personal display" and investigate the degree of citation personal display between physicists, an examination of the factors will be placed in our subsequent study. We believe that it is significant to seek factors that influence citation personal display, and that such investigation will also enrich studies of related topics.

Finally, we want to point that the so-called "citation personal display" can be regarded as the citation recognition or citation preference, in that a researcher may know of citation indices but prefer to ignore them. Personal preference of researchers is not tested in this work and can only be inferred from their behavior, and a further survey needs to be constructed. Another thing of note is that the sample size of this study is relatively small, which may cause a skewing of results. Whereas, statistically speaking, the sample size of this investigation is enough to describe the phenomenon that physicists in question have different degrees of citation personal display. Of course our results will be more persuasive with an increase of sample size, and this will be remedied in our next research.

Another thing we have to point out is that this study has considered only physicists, which greatly limits the generalizability of the findings. More subjects will be included in follow-up study. Do other disciplineshave consistent figures for citation personal display? This seems to be a worthy research direction. Further, the results of our study show that there is still a low proportion of citation personal display (14.8%) among our discussed researchers in physics. It would be interesting to know how the rate of citation personal display changes over time, which could be a valuable research topic.

**Acknowledgments**

This research was supported by the National Natural Science Foundation of China (Grant No. 71273250).

**Appendix**

Table V    List of universities studied

| University abbreviation | University name |
|---|---|
| Harvard | Harvard University |
| MIT | Massachusetts Institute of Technology |
| Stanford | Stanford University |
| Caltech | California Institute of Technology |
| Princeton | Princeton University |
| Oxon. | University of Oxford |
| Cambridge | University of Cambridge |
| ICL | Imperial College London |
| Peking | Peking University |
| Tsinghua | Tsinghua University |
| Fudan | Fudan University |